\documentclass{article}
\usepackage{ijcai19}

\usepackage{times}
\usepackage{soul}
\usepackage{url}
\usepackage[hidelinks]{hyperref}
\usepackage[utf8]{inputenc}
\usepackage[small]{caption}
\usepackage{graphicx}
\usepackage{amsmath}
\usepackage{booktabs}
\usepackage{xcolor}
\title{Building Bridges: Generative Artworks to Explore AI Ethics}

\author{Ramya Srinivasan$^{1}$ and Devi Parikh$^{2}$\\
 $^{1}$Fujitsu Research of America\\
 $^{2}$Georgia Tech and Facebook AI Research\\
 }

\begin{document}

\maketitle




\begin{abstract}
In recent years, there has been an increased emphasis on understanding and mitigating adverse impacts of artificial intelligence (AI) technologies on society. Across academia, industry, and government bodies, a variety of endeavours are being pursued towards enhancing AI ethics. A significant challenge in the design of ethical AI systems is that there are multiple stakeholders in the AI pipeline, each with their own set of 
constraints and interests. 
These different perspectives are often not understood, due in part to communication gaps.
For example, AI researchers who design and develop AI models are not necessarily aware of the instability induced in consumers' lives by the compounded effects of AI decisions. Educating different stakeholders about their roles and responsibilities in the broader context becomes necessary. 
In this position paper, we outline some potential ways in which generative artworks can play this role by serving as accessible and powerful educational tools for surfacing different perspectives. We hope to spark interdisciplinary discussions about computational creativity broadly as a tool for enhancing AI ethics. 
\end{abstract}

\section{Introduction}
From finance and healthcare to education and policy making, artificial intelligence (AI) technologies are being employed across a variety of domains to make important decisions about people's lives. Amidst this growing prevalence, concerns have been raised about the bias and discrimination associated with such automated decisions \cite{joy,ziad,kristian}. Indeed, numerous research efforts are centered around AI ethics, including bias mitigation \cite{manish}, analyzing impacts of AI decisions \cite{lydia}, guidelines for datasets \cite{kate}, incorporating ethics in AI system's decision making \cite{aditya,nick}, and explainable AI \cite{russell}. Significant gaps remain in transforming AI technologies into large scale systems of ethical value.

A significant challenge in the design of ethical AI systems is the fact that there are multiple stakeholders in the AI pipeline, each with their own set of requirements and responsibilities.  AI researchers tend to focus on designing algorithms that can achieve high accuracy and explain the rationale behind the decision, developers audit the system and analyze its failure modes to enhance robustness and safety, decision makers need to justify the use of these systems by examining if the decisions can be trusted, consumers not only seek systems that are trust-worthy, safe, and explainable, but also allow for redressing adverse decisions. 

The subjective perspectives of various stakeholders are seldom addressed jointly due to communication gaps and contradicting requirements. By and large, a vast majority of the consumers do not understand how the AI decisions are made, and if the objectives of the AI aligns with the values they care about. AI researchers do not necessarily understand the compounded adverse effects of AI decisions on consumers. Such communication gaps hinder the design, development, and adoption of ethical AI systems. 

Participatory design is often seen as a way to help AI researchers and developers analyze the adverse impacts of these technologies on society. For example, the authors in \cite{fish} advocate for the use of reflexive design  whereby a researcher has to introspect on their own role with respect to other stakeholders while designing the AI system. In \cite{barbosa}, the authors suggest a language based epistemic tool that uses principles from semiotic engineering to help bridge gaps between designers and users of AI systems. 

In this regard, artworks of all forms -- analog or digital, designed or generative, rule based or learning based -- can serve as powerful visualization tools to educate, platforms to gather opinions of, and facilitators 
of empathy across different stakeholders. As the adage {\it `A picture is worth a thousand words'} states, visuals are more effective than verbal modes of communication, and can make concepts accessible that otherwise may require specialized knowledge to grasp. 

While there is benefit from all forms of artworks, in this position paper, we focus on the potential role of generative artworks in enhancing AI ethics. A motivation for such a focus stems from the work of \cite{rediet}, wherein the authors highlight the various roles of computational research in general to bring about social change. 

In particular, we outline four pathways through which generative artworks can help bridge the communication gap across stakeholders in the AI pipeline. 
First, generative artworks can elucidate different subjective viewpoints by showcasing various ethical perspectives pertaining to a situation. Second, generative artworks can help AI researchers and developers visualize counterfactual situations that can help them in understanding some adverse effects of AI decisions on individuals. Third, generative artworks can help in understanding misalignment in the AI pipeline such as those that exist between constructs (e.g. fairness) and their measurements (e.g. demographic parity). Finally, generative artworks can shed light on non-western perspectives that are seldom considered in AI system design and evaluation. More broadly, we hope the ideas outlined in the paper spark interdisciplinary discussions related to the ways in which computational creativity can be leveraged towards enhancing AI ethics. Sections 2-5 elaborate on these broad ideas, and Section 6 concludes the paper. Note that as with any computational tool, there can be downsides too. As the authors in \cite{srinivasan,vinay} note, generative artworks can be embedded with different types of biases. Thus, it is necessary to employ these tools mindfully. 

\section{Visualizing different ethical perspectives}
Ethical theories have long been perceived as foundations for decision making \cite{pratt}. There are many ethical theories such as deontological ethics, utilitarian ethics, consequentialism, etc. The suitability and adequacy of a theory can vary based on the context and issue under consideration \cite{sabine}. Context could include factors such as who is seeking the decision, who is making the decision, who is affected by the decision, what is the use-case, and so on. 

Each ethical theory emphasizes specific principles in decision making, and can thus shed light on varying viewpoints relevant in a given context. For example, in utilitarian ethics, the emphasis is on maximizing the well-being of all stakeholders. In deontological ethics, the emphasis is on following the laws and regulations. In virtue ethics, the emphasis is on moral values. In subjectivism, the emphasis is on individual beliefs, experiences, and opinions. In feminist ethics, the emphasis is on interpersonal aspects such as caring, interdependence, and the ethical requirements of particular relationships. Situational ethics takes into account the particular context of an act, rather than judging it according to absolute moral standards or normative rules. Consequentialism focuses on the consequences of an act.

Thus, even within the context of a single problem setting, there can be diverse viewpoints about what is right, fair, just, or appropriate. 
In order to enhance AI ethics, it thus becomes important to educate AI researchers and developers about these diverse viewpoints and thereby aid in reflexive design. There are some works in which researchers have leveraged artworks to illustrate issues pertaining to differnt ethical theories. For example, American artist Rashaad Newsome tells stories of racial injustice through the lens of colonialism to highlight existing shortcomings of AI models. In \cite{falaah}, opinions about fairness notions such as equality of opportunity have been illustrated via comics by artist Falaah Arif Khan. Inspired by the Theory of New Aesthetics \cite{leech}, in \cite{meshi},  artist Avital Meshi deconstructs the notion of ``whiteness" in face recognition algorithms by using performance as a tool. The artist examines the behavior of the algorithm by performing/playing different expressions to observe the algorithm's confidence in recognizing the artist's race.  In {\it ImageNet Roulette}, an art project \cite{paglen}, biases associated with large scale image datasets and their adverse consequences are highlighted. Designed by Microsoft researcher Kate Crawford and artist Trevor Paglen, ImageNet Roulette was on view in an exhibition called “Training Humans,” at the Fondazione Prada’s Osservatorio space, Milan. Unlike these works which focus on a broad range of artworks, our focus here is on analyzing the potential of {\it generative} artworks in elucidating different ethical perspectives. 



Generative artworks can be powerful visualization tools to aid AI researchers and developers in understanding diverse ethical perspectives. For example, through generative art techniques, it may be possible to visualize the compounded effects of adverse AI decisions and the instability induced in people's lives, a consequentialism ethical perspective which is otherwise hard to infer. Similarly, it may be possible to generate visualizations corresponding to different subjective opinions to better understand the differences. For example, for some people, an AI decision going one way or the other could cause a lot of chaos. Others may stay relatively unaffected. Through generative artworks, it might be possible to show how based on certain demographic and cultural characteristics of the person, perceptions can vary.
\section{Visualizing counterfactuals}
Generative artworks could also aid in visualizing counterfactual situations which in turn can be beneficial in reflexive design via empathy fostering. Counterfactual thinking is a concept in psychology that involves the human tendency to create possible alternatives to life events that have already occurred, i.e. something that is contrary to the given facts \cite{kai}. 

Counterfactual thinking can help in engendering empathy by enabling one to visualize situations through another person's world. Thus, certain situations that may be irrelevant in one person's context, but relevant in another person's context, can be understood via such counterfactual visualizations. In particular, counterfactual visualizations can aid in understanding emotions of people beyond oneself, to understand their pain points, and anticipate their reactions. Thus, such visualizations can help in creating a mental symbiosis between developers and users of AI technologies. 


There has been some work related to empathy fostering in AI. For example, in a project titled ``Deep Empathy", researchers utilized deep learning techniques to learn characteristics of Syrian neighborhoods after the war, and used these features to transform images of cities all over the world, simulating how they would look if they suffered disasters like those in Syria \cite{free}. In the words of the researchers, ``Deep Empathy gets you closer to the realities of those that suffer the most, by helping you imagine what neighbourhoods around the world would look like if hit by a disaster.'' \cite{mit}.

In a similar vein, generative artworks could be used as tools to visualize the consequences of AI decisions so AI researchers and developers (for instance), who may not necessarily be affected by the decision, can empathize with the impacted population. 

\section{Visualizing mismatches in the AI pipeline}
More often than not, computational systems involve quantitatively modeling abstract concepts or constructs which may or may not be observable. AI systems are no exceptions in this regard. For example, consider the construct of ``fairness".  Fairness is essentially a contested construct with multiple context dependent and sometimes contradicting theoretical understandings \cite{jacobs}. This makes it inherently difficult to quantify or measure fairness. Any quantification or measurement process necessarily makes assumptions which may introduce misalignment between the theoretical understanding and its measurements. Many adverse effects of AI systems can be seen as direct consequences of such mismatches \cite{jacobs}. 

Furthermore, there may be unobservable factors that affect the constructs themselves. Consider for example a construct such as ``skill" or ``ability" which is relevant across many applications such as hiring and admissions.  As the authors in \cite{corrina} note, these constructs can be influenced by both innate potential specific to the individual and other factors such as socio-economic status. Thus, a mismatch can be introduced even before measuring a construct. 

Generative artworks could aid in visualizing such mismatches. For example, it may be possible to highlight differences in measurement of similar constructs, thereby aiding AI researchers and developers in understanding system behavior. Consider an AI based hiring use case as an example. Suppose one of the features in making the decision concerns measuring social skills of the candidate. In this regard, one might expect the constructs ``self-esteem" and ``confidence" to be related. 
Visualizations of AI system's behavior under different scenarios could reveal whether it treats these constructs similarly -- whether it exhibits ``convergent validity" \cite{jacobs}. Convergent validity refers to the degree to which two measures of constructs that theoretically should be related, are in fact related. 

Generative artworks could also be used to showcase the influence of factors that are not considered in construct formulation, thereby shedding light on inductive biases. For example, in using the fairness notion of demographic parity, AI researchers may have hypothesized that the measured abilities are true indicators of one's skills. However, due to certain unjust circumstances, certain individuals may have been inhibited from acquiring the required skills. An AI system that overlooks such broader contextual information is prone to bias. By showcasing the mismatches between measured abilities and innate abilities, it is possible to help AI researchers understand that although two individuals can have different measured abilities, they could have the same innate abilities, thus helping them in examining their hypothesis.  

\begin{figure}[t]
  \centering
    \includegraphics[width=0.5\textwidth]{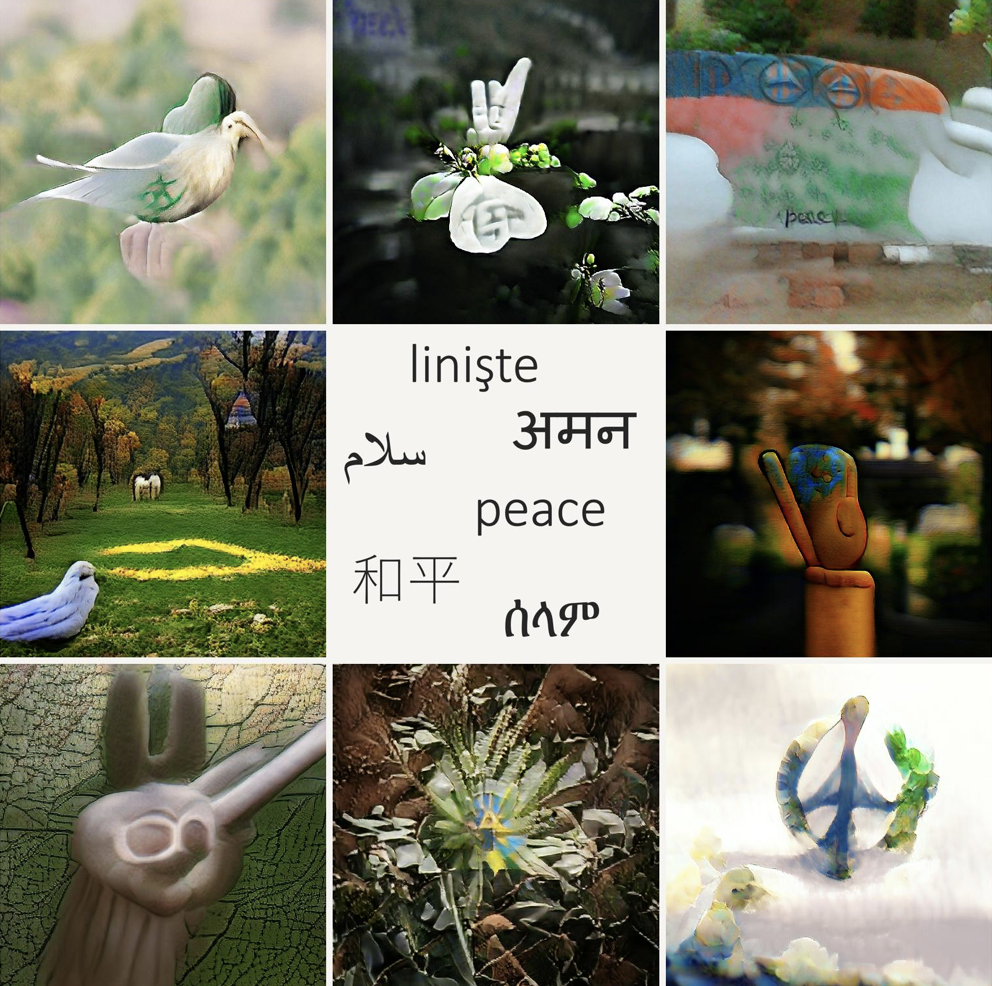}
    \caption{An illustration of the concept of ``peace" in different countries and languages as generated by an AI model. Image source \url{https://www.cc.gatech.edu/~parikh/horizons.html}}
    \label{fig:horizon}
\end{figure}

\section{Visualizing non-western perspectives}
Most works concerning AI ethics are influenced heavily by western cultures in their definitions, axes of discrimination, and philosophical roots \cite{nithya}. For example, a majority of fairness research study race and gender based discrimination, which are dominant in the American settings. Axes of discrimination relevant to other geographies and cultures such as societal hierarchies and economic inequalities are seldom considered. 

It becomes important to consider notions beyond fairness such as those concerning restorative justice and those based on moral foundations as these can help shed light on philosophical viewpoints that are relevant in local cultural and geographical contexts \cite{nithya}.
In this regard, generative artworks can serve as visualizations of social, cultural, and economic differences that exist across geographies. For example, through generative artworks it may be possible to highlight different viewpoints regarding fairness based on the local context such as social practices, religious beliefs, economic status, etc.  By training generative models on data across cultures and looking at the latent visualizations, it might also be possible to view how different everyday practices (e.g. dress, food, etc.) and objects (e.g. furniture, houses, etc.) can vary across cultures thereby shedding light on local contexts which can be valuable in AI system design. An example is shown in Figure \ref{fig:horizon}. Curiosity to understand why the model had the specific interpretations of ``peace'' in different contexts (languages, countries) enabled conversations between individuals from around the world, who may not have otherwise connected~\cite{parikh}.  

\section{Conclusions}
We posit that generative artworks can help in narrowing communication gaps across various stakeholders in the AI pipeline by serving as powerful and accessible educational tools to visualize different perspectives, thereby helping in enhancing AI ethics. In particular, we outlined how generative artworks can aid in the visualization of diverse ethical viewpoints, counterfactual scenarios, mismatches in the AI pipeline, and non-western perspectives. We hope our work sparks interdisciplinary discussions related to the use of computational creativity towards enhancing AI ethics.

\bibliographystyle{named}
\bibliography{ijcai19}

\begin{thebibliography}{}

\bibitem[\protect\citeauthoryear{Abebe \bgroup \em et al.\egroup
  }{2020}]{rediet}
Rediet Abebe, Solon Barocas, Jon Kleinberg, Karen Levy, Manish Raghavan, and
  David~G. Robinson.
\newblock {Roles for Computing in Social Change}.
\newblock {\em FAccT}, 2020.

\bibitem[\protect\citeauthoryear{Barbosa \bgroup \em et al.\egroup
  }{2021}]{barbosa}
Simone Diniz~Junqueira Barbosa, Gabriel Diniz~Junqueira Barbosa,
  Clarisse~Sieckenius de~Souza, and Carla~Faria Leitao.
\newblock {A Semiotics-based Epistemic Tool to Reason about Ethical Issues in
  Digital Technology Design and Development}.
\newblock {\em FAccT}, 2021.

\bibitem[\protect\citeauthoryear{Bostrom and Yudkowsky}{2018}]{nick}
Nick Bostrom and Eliezer Yudkowsky.
\newblock {The Ethics of Artificial Intelligence}.
\newblock {\em Cambridge Handbook of Artificial Intelligence}, 2018.

\bibitem[\protect\citeauthoryear{Buolamwini and Gebru}{2018}]{joy}
Joy Buolamwini and Timnit Gebru.
\newblock {Gender Shades: Intersectional Accuracy Disparities in Commercial
  Gender Classification}.
\newblock {\em FAccT}, 2018.

\bibitem[\protect\citeauthoryear{Crawford and Paglen}{2019}]{paglen}
Kate Crawford and Trevor Paglen.
\newblock {Excavating AI: The Politics of Images in Machine Learning Training
  Sets}.
\newblock {\em https://www.excavating.ai}, 2019.

\bibitem[\protect\citeauthoryear{DeepEmpathy}{2021}]{mit}
DeepEmpathy.
\newblock {What Would Your City Look Like After a Disaster?}
\newblock {\em https://www.media.mit.edu/projects/deep-empathy/overview/},
  2021.

\bibitem[\protect\citeauthoryear{Epstude and Roese}{2008}]{kai}
Kai Epstude and Neal~J. Roese.
\newblock {The Functional Theory of Counterfactual Thinking}.
\newblock {\em Personality and Social Psychology Review}, 2008.

\bibitem[\protect\citeauthoryear{Fish and Stark}{2021}]{fish}
Benjamin Fish and Luke Stark.
\newblock {Reflexive Design for Fairness and Other Human Values in Formal
  Models}.
\newblock {\em AIES}, 2021.

\bibitem[\protect\citeauthoryear{Free}{2018}]{free}
Lynsee Free.
\newblock {Can AI Help Humans Be More Empathetic?}
\newblock {\em https://strelkamag.com/en/article/ai-mit-unicef-deep-empathy},
  2018.

\bibitem[\protect\citeauthoryear{Gebru \bgroup \em et al.\egroup }{2018}]{kate}
Timnit Gebru, Jamie Morgenstern, Briana Vecchione, Jennifer~Wortman Vaughan,
  Hanna Wallach, Hal~Daumeé III, and Kate Crawford.
\newblock {Datasheets for Datasets}.
\newblock {\em Workshop on Fairness, Accountability, and Transparency in
  Machine Learning}, 2018.

\bibitem[\protect\citeauthoryear{Hegde \bgroup \em et al.\egroup
  }{2020}]{aditya}
Aditya Hegde, Vibhav Agarwal, and Shrisha Rao.
\newblock {Ethics, Prosperity and Society: Moral Evaluation Using Virtue Ethics
  And Utilitarianism}.
\newblock {\em IJCAI}, 2020.

\bibitem[\protect\citeauthoryear{Hertweck \bgroup \em et al.\egroup
  }{2021}]{corrina}
Corrina Hertweck, Christoph Heitz, and Michele Loi.
\newblock {On the Moral Justification of Statistical Parity}.
\newblock {\em FAccT}, 2021.

\bibitem[\protect\citeauthoryear{Jacobs and Wallach}{2021}]{jacobs}
Abigail~Z. Jacobs and Hanna Wallach.
\newblock {Measurement and Fairness}.
\newblock {\em FAccT}, 2021.

\bibitem[\protect\citeauthoryear{Khan \bgroup \em et al.\egroup
  }{2021}]{falaah}
Falaah~Arif Khan, Eleni Mannis, and Julia Stoyanovich.
\newblock {Translation Tutorial: Fairness and Friends}.
\newblock {\em FAccT}, 2021.

\bibitem[\protect\citeauthoryear{Kurutach \bgroup \em et al.\egroup
  }{2018}]{russell}
Thanard Kurutach, Aviv Tamar, Ge~Yang, Stuart Russell, and Pieter Abbeel.
\newblock {Learning Plannable Representations with Causal InfoGAN}.
\newblock {\em NeurIPS}, 2018.

\bibitem[\protect\citeauthoryear{Leech}{2011}]{leech}
Guy Leech.
\newblock {James Bridle – Waving at the Machines}.
\newblock {\em
  https://webdirections.org/resources/james-bridle-waving-at-the-machines/},
  2011.

\bibitem[\protect\citeauthoryear{Liu \bgroup \em et al.\egroup }{2018}]{lydia}
Lydia Liu, Sarah Dean, Esther Rolf, Max Simchowitz, and Moritz Hardt.
\newblock {Delayed Impact of Fair Machine Learning }.
\newblock {\em ICML}, 2018.

\bibitem[\protect\citeauthoryear{Lum \bgroup \em et al.\egroup
  }{2020}]{kristian}
Kristian Lum, Chesa Boudin, and Megan Price.
\newblock {The Impact of Overbooking on a Pre-trial Risk Assessment Tool}.
\newblock {\em FAccT}, 2020.

\bibitem[\protect\citeauthoryear{Meshi}{2021}]{meshi}
Avital Meshi.
\newblock {Deconstructing Whiteness}.
\newblock {\em CVPR Workshop on Ethical Considerations in Creative Applications
  of Computer Vision}, 2021.

\bibitem[\protect\citeauthoryear{Obermeyer \bgroup \em et al.\egroup
  }{2019}]{ziad}
Ziad Obermeyer, Brian Powers, Christine Vogeli, and Sendhil Mullainathan.
\newblock {Dissecting Racial Bias in an Algorithm Used to Manage the Health of
  Populations}.
\newblock {\em Science}, 2019.

\bibitem[\protect\citeauthoryear{Parikh \bgroup \em et al.\egroup
  }{2021}]{parikh}
Devi Parikh, Amal Alabdulkarim, Gemmechu Hassena, Oana Ignat, Jiasen Lu, and
  Ryan Murdock.
\newblock {Expanding Horizons: Visual Indeterminacy as A Vehicle For Inquiry
  and New Connections}.
\newblock {\em https://www.cc.gatech.edu/~parikh/horizons.html}, 2021.

\bibitem[\protect\citeauthoryear{Prabhu and X}{2011}]{vinay}
Vinay~Uday Prabhu and Isiain X.
\newblock {A taxonomy of concerns concerning neural art}.
\newblock {\em CVPR Workshop on Ethical Considerations in Creative Applications
  of Computer Vision}, 2011.

\bibitem[\protect\citeauthoryear{Pratt}{1994}]{pratt}
Cornelius~B. Pratt.
\newblock {Applying Classical Ethical Theories to Ethical Decision Making in
  Public Relations: Perrier's Product Recall}.
\newblock {\em Management Communication Quarterly}, 1994.

\bibitem[\protect\citeauthoryear{Raghavan \bgroup \em et al.\egroup
  }{2020}]{manish}
Manish Raghavan, Solon Barocas, Jon Kleinberg, and Karen Levy.
\newblock {Mitigating Bias in Algorithmic Hiring: Evaluating Claims and
  Practices}.
\newblock {\em FAccT}, 2020.

\bibitem[\protect\citeauthoryear{Salloch \bgroup \em et al.\egroup
  }{2015}]{sabine}
Sabine Salloch, Sebastian Wäscher, Jochen Vollmann, and Jan Schildmann.
\newblock {The Normative Background of Empirical-ethical Research: First Steps
  Towards a Transparent and Reasoned Approach in the Selection of an Ethical
  Theory}.
\newblock {\em BMC Med Ethics}, 2015.

\bibitem[\protect\citeauthoryear{Sambasivan \bgroup \em et al.\egroup
  }{2021}]{nithya}
Nithya Sambasivan, Erin Arnesen, Ben Hutchinson, Tulsee Doshi, and Vinodkumar
  Prabhakaran.
\newblock {Re-imagining Algorithmic Fairness in India and Beyond}.
\newblock {\em FAccT}, 2021.

\bibitem[\protect\citeauthoryear{Srinivasan and Uchino}{2021}]{srinivasan}
Ramya Srinivasan and Kanji Uchino.
\newblock {Biases in Generative Arts---A Causal Look from the Lens of Art
  History}.
\newblock {\em FAccT}, 2021.

\end{thebibliography}
\end{document}